# Observation of topological superconductivity on the surface of an iron-based superconductor

Peng Zhang[1*], Koichiro Yaji[1], Takahiro Hashimoto[1], Yuichi Ota[1], Takeshi Kondo[1], Kozo Okazaki[1], Zhijun Wang[2], Jinsheng Wen[3], G. D. Gu[4], Hong Ding[5*], and Shik Shin[1*]

[1]Institute for Solid State Physics, University of Tokyo, Kashiwa, Chiba 277-8581, Japan

[2]Department of Physics, Princeton University, Princeton, New Jersey 08544, USA

[3]National Laboratory of Solid State Microstructures and Department of Physics, Nanjing University, Nanjing 210093, China

[4]Condensed Matter Physics and Materials Science Department, Brookhaven National Laboratory, Upton, New York 11973, USA

[5]Beijing National Laboratory for Condensed Matter Physics, and Institute of Physics, Chinese Academy of Sciences, Beijing 100190, China

*Correspondence to: zhangpeng@issp.u-tokyo.ac.jp, dingh@iphy.ac.cn, shin@issp.u-tokyo.ac.jp

**Topological superconductors, whose edge hosts Majorana bound states or Majorana fermions that obey non-Abelian statistics, can be used for low-decoherence quantum computations. Most of the proposed topological superconductors are realized with spin-helical states through proximity effect to BCS superconductors. However, such approaches are difficult for further studies and applications because of the low transition temperatures and complicated hetero-structures. Here by using high-resolution spin-resolved and angle-resolved photoelectron spectroscopy, we discover that the iron-based superconductor $FeTe_{1-x}Se_x$ (x = 0.45, Tc = 14.5 K) hosts Dirac-cone type spin-helical surface states at Fermi level, which open an s-wave SC gap below Tc. Our study proves that the surface states of $FeTe_{0.55}Se_{0.45}$ are 2D topologically superconducting, and thus provides a simple and possibly high-Tc platform for realizing Majorana fermions.**

A topological superconductor (TSC) has a non-trivial topology when opening a superconducting (SC) gap, which is associated with the emergence of zero energy excitations that are their own antiparticles (*1,2*). These zero-energy bound states are generally called Majorana bound states or Majorana fermions. Their potential applications in quantum computations inspire an intensive search for topological superconductors and Majorana fermions. Typically, there are two ways to realize topological superconductivity. One is to realize a p-wave superconductor, whose SC state itself is topologically non-trivial. Thus it is an intrinsic topological superconductor. The prominent candidates are $Sr_2RuO_4$ and $Cu_xBi_2Se_3$. However, p-wave superconductivity is very sensitive to disorder and the experimental confirmation of the topological edge states in these systems is still elusive, and any application is highly challenging (*3-5*). Another way is to realize s-wave superconductivity on spin-helical states (*6*). This can be in general realized by a topological insulator in proximity to a BCS superconductor, or a semiconductor with Rashba spin-split states in a magnetic field and in proximity to a BCS superconductor, and some of the designs have yielded strong experimental evidences of Majorana fermions (*7-11*). However, this approach generally requires a long SC coherence length which in principle prohibits applications of high-Tc superconductors. Furthermore, the complicated hetero-structures are obstacles for research and applications. In this work, we discover that $FeTe_{0.55}Se_{0.45}$, which is one of the high-Tc Fe-based superconductors and can have relatively high Tc under certain conditions, hosts topological superconducting states on the surface, in accordance with our theoretical predictions (*12,13*). This intrinsic topological superconductor, which takes advantage of the natural surface and interband SC coherence in the momentum space, can overcome all the disadvantages mentioned above, and thus pave a new and exciting route for realizing topological superconductivity and Majorana fermions under higher temperature.

**First-principles calculations**

Fe(Te,Se) has the simplest crystal structure among Fe-based superconductors, as shown in Fig.1A, making it easy to get high quality single crystals and thin films. Its Tc can reach ~ 30 K under pressure (*14*), and even above 40 K with monolayer thin film (*15*). Its in-plane electronic structure is similar to most of the iron-based superconductors: There are two hole-like Fermi surfaces at Brillouin zone (BZ) center ($\Gamma$) and two electron-like FSs at BZ corner (M), as shown in Fig.1B. Correspondingly, for a cut along $\Gamma M$, there are three hole-like bands (Two of them crossing Fermi level ($E_F$)) at $\Gamma$ and two electron-like bands at M. One novelty for this material is that, when considering the out-of-plane electronic structure, band calculations predict that $FeTe_{0.5}Se_{0.5}$ has a non-trivial topological invariance and hosts topological surface states near $E_F$ (*12,13,16*).

Calculations show that the topological order originates from the Te substitution, which not only introduces large spin-orbit coupling (SOC) (*17*), but also shifts $p_z$ band downward to $E_F$ (*12*), whereas the $p_z$ band in FeSe or iron pnictides is generally above $E_F$ (*18,19*). These facts make Fe(Te,Se) unique on topological properties. Fig.1D shows the band structure along $\Gamma M$ and $\Gamma Z$. We notice that along $\Gamma Z$ $p_z$ band has a large dispersion and crosses $d_{xz}$ band near $E_F$. At the crossing point, an SOC gap opens. Further analysis shows that $p_z$ band has a "-" parity sign for the inversion symmetry, while $d_{xz}$ band has a "+" sign. With these necessary ingredients, the calculated non-trivial topological invariance confirms that $FeTe_{0.5}Se_{0.5}$ indeed hosts strong topological surface states (TSS) near $E_F$ (*12*). To show the predicted TSS clearly, we project the band structure onto the (001) surface in Fig.1E. The Dirac-cone type TSS are located just near $E_F$, inside the SOC gap of bulk valence band and bulk conduction band. When $FeTe_{0.5}Se_{0.5}$ enters SC state with s-wave gaps, the TSS can be induced to be superconducting, as shown in Fig.1F. The spin-helical and s-wave SC characters together would make the surface states topologically superconducting (*6*).

**Dirac-cone type spin-helical surface band and s-wave superconducting gap**

To experimentally prove that $FeTe_xSe_{1-x}$ (x ~ 0.5) is a topological superconductor with intrinsic topological surface states and s-wave superconductivity on the surface, one needs to observe the following three phenomena in spectroscopic measurements: 1) Dirac-cone type surface states, 2) spin-helical texture of the surface states, 3) s-wave SC gap of the surface states when T < Tc. Previously we obtained some experimental evidences for the band inversion of the bulk $p_z$ and $d_{xz}$ bands (*12,20*). However, the topological surface band is never directly observed, due to the small energy and momentum scales. The SOC gap is estimated to be about 10 meV in the calculations, which makes it extremely difficult to resolve the Dirac-cone type surface states in angle-resolved photoelectron spectroscopy (ARPES). In the previous ARPES experiments, only the three $t_{2g}$ ($d_{xy}$, $d_{yz}$ and $d_{xz}$) and $p_z$ bulk bands are observed at Γ (*12,21,22*). In the current experiment present below, by using high energy and momentum resolution ARPES (HR-ARPES) (Energy resolution ~ 1.4 meV) (*23*) and spin-resolved ARPES (SARPES) (Energy resolution ~ 5.5 meV) (*24*), we are able to observe the three necessary phenomena required for topological superconductivity in high-quality single crystals $FeTe_{0.55}Se_{0.45}$.

We first demonstrate the observation of the Dirac-cone type surface states. Fig.2, A and C show the high resolution cuts of the band structure around Γ with p and s-polarized photons, respectively. According to the matrix element effect (Supplementary Materials Part I), both the surface and the bulk bands ($p_z$ and $d_{xz}$) should be visible for p-polarized photons, while only the bulk valence band ($d_{xz}$) is visible for s-polarized photons. The momentum distribution curve (MDC) curvature plot (an improved version of second derivative method) (*25*) of the data with p-polarized photons shows a clear Dirac-cone type band, as displayed in Fig.2B. Meanwhile, we get a good parabolic band by fitting the energy distribution curve (EDC) peaks of the data with s-polarized photons (Fig.2C). Combining the bands observed in Fig.2, A to C, we conclude that the Dirac-cone type band (Blue lines in Fig.2B) is the topological surface band, and the parabolic band (White curve in Fig.2B or red curve in Fig.2C) is the bulk valence band. Furthermore, we directly separate the bulk valence band from the Dirac-cone type surface band with the data at very low temperature (2.4 K) when the spectral features are narrower, as shown in Fig.2, D and E. We overlap the Dirac-cone type surface band in Fig.2B and the parabolic bulk band in Fig.2C onto the EDC curvature plot in Fig.2E. The extracted bands overlap well with the curvature intensity plot, confirming the existence of the parabolic bulk band and the Dirac-cone type surface band. The overall band structure is summarized in Fig.2F.

Next we carry out high resolution spin-resolved experiments to check the spin polarization of the Dirac-cone type band. Two EDCs at cuts indicated in Fig.3A were measured. If the Dirac-cone type band indeed comes from the spin-polarized surface states, the EDCs at Cut1 and Cut2 should show reversed spin polarizations. Indeed, the spin-resolved EDCs in Fig.3, B and D show that the spin polarizations are reversed for Cut1 and Cut2, while the background shows no spin polarization (Fig.3, C and E). These data clearly confirm the spin-helical texture which is the direct consequence of "spin-momentum locking" of topological surface states. We also measured another two EDCs at different positions on the Fermi surface (FS) (Supplementary Materials Part III). The spin polarizations of all the four EDCs are consistent with the spin-helical texture predicted by the theory (*12*). We note that the small spin polarizations in Fig.3, C and E could partly come from the large broadening of the SARPES data, originating from the lower resolution of the SARPES, as shown in Fig. 3F.

At last, we show there is an s-wave gap opening for the topological surface band. Fig.4A displays the evolution of one EDC from the surface band with temperature. The SC coherence peak gradually builds up with decreasing temperature. In another aspect, the symmetrized EDCs in Fig.4B show the gap closing above Tc clearly. The relation between the SC gap size and temperature is displayed in Fig.4C, in good consistency with BCS theory. The EDC divided by the corresponding Fermi function (Fig.4C inset) shows a clear peak at the symmetric position above $E_F$, which comes from the particle-hole mixing of the Bogolubov quasiparticles, and thus proving the superconducting nature of the coherence peak. The momentum dependent measurement of the SC gap size shows no anisotropy, as displayed in Fig.4, D and E, consistent with the s-wave superconducting nature of iron-based superconductors (*26-28*). The gap size of the surface band is about 1.8 meV, which is smaller than the bulk gap size of 2.5 meV for the hole band and 4.2 meV for the electron band, as reported in References (*26,27*). This result is consistent with induced superconductivity on the surface, and may even suggest that the induced superconductivity mainly comes from interband scattering from the neighboring hole-like band.

**Majorana fermions and bound states on the edge**

We summarize our results in Fig.5A. There exists a Dirac-cone type topological surface band on the surface of $FeTe_{0.55}Se_{0.45}$. When the bulk bands open s-wave gaps, the surface band is induced to be superconducting through interband scattering. Because of the spin-helical texture, the surface enters topologically superconducting states, while the bulk superconductivity is topologically trivial. When applying an external magnetic field, there will be a pair of Majorana bound states in the vortices, locating at the two ends of the vortices, as shown in Fig.5B. Sharp zero bias peaks have been already observed in the vortices in recent STM experiment (*29*). These zero bias peaks are in general absent in many other iron-based superconductors, and thus likely to be the Majorana bound states according to our results. There is also a report of a robust zero bias peak at excess Fe sites, which may also be related to topologically superconducting states (*30*). Furthermore, if a magnetic domain is deposited on the surface which kills the superconductivity in the domain, there should be Majorana fermions along the boundary. As a result of the intrinsic topological superconductivity on the natural surface, it is fairly easy to produce Majorana bound states and fermions. Because of the relatively high Tc and the easiness to grow high-quality single crystals and thin films, Fe(Te,Se) will be a perfect platform to study Majorana bound states, and may further advance the research on quantum computations.

**References and notes:**
1. X.-L. Qi, S.-C. Zhang, Rev. Mod. Phys. **83**, 1057 (2011).
2. C. Nayak, S. H. Simon, A. Stern, M. Freedman, S. Das Sarma, Rev. Mod. Phys. **80**, 1083 (2008).
3. L. Fu, E. Berg, Phys. Rev. Lett. **105**, 097001 (2010).
4. S. Sasaki et al., Phys. Rev. Lett. **107**, 217001 (2011).
5. N. Levy et al., Phys. Rev. Lett. **110**, 117001 (2013).
6. L. Fu, C. L. Kane, Phys. Rev. Lett. **100**, 096407 (2008).
7. V. Mourik et al., Science **336**, 1003 (2012).
8. S. Nadj-Perge et al., Science **346**, 602 (2014).
9. S. M. Albrecht et al., Nature **531**, 206 (2016).
10. S.-Y. Xu et al., Nat. Phys. **10**, 943 (2014).
11. H.-H. Sun et al., Phys. Rev. Lett. **116**, 257003 (2016).
12. Z. Wang et al., Phys. Rev. B **92**, 115119 (2015).

13. X. Wu, S. Qin, Y. Liang, H. Fan, J. Hu, Phys. Rev. B **93**, 115129 (2016).
14. K. Horigane, N. Takeshita, C.-H. Lee, H. Hiraka, K. Yamada, J. Phys. Soc. Jpn. **78**, 063705 (2009).
15. F. Li et al., Phys. Rev. B **91**, 220503 (2015).
16. G. Xu, B. Lian, P. Tang, X.-L. Qi, S.-C. Zhang, Phys. Rev. Lett. **117**, 047001 (2016).
17. P. D. Johnson et al., Phys. Rev. Lett. **114**, 167001 (2015).
18. S. Graser et al., Phys. Rev. B **81**, 214503 (2010).
19. H. Eschrig, A. Lankau, K. Koepernik, Phys. Rev. B **81**, 155447 (2010).
20. X. Shi et al., Sci. Bull. **62**, 503 (2017).
21. Y. Lubashevsky, E. Lahoud, K. Chashka, D. Podolsky, A. Kanigel, Nat. Phys. **8**, 309 (2012).
22. P. Zhang et al., Appl. Phys. Lett. **105**, 172601 (2014).
23. K. Okazaki et al., Science **337**, 1314 (2012).
24. K. Yaji et al., Rev. Sci. Instrum. **87**, 053111 (2016).
25. P. Zhang et al., Rev. Sci. Instrum. **82**, 043712 (2011).
26. H. Miao et al., Phys. Rev. B **85**, 094506 (2012).
27. K. Okazaki et al., Sci. Rep. **4**, 4109 (2014).
28. D. C. Johnston, Adv. Phys. **59**, 803 (2010).
29. F. Massee et al., Sci. Adv. **1**, e1500033 (2015).
30. J.-X. Yin et al., Nat. Phys. **11**, 543 (2015).

**Acknowledgments:**

We acknowledge C. Bareille, Y. Ishida, K. Kuroda, R. Noguchi, A. Tsuzuki for assistance in the experiments and J. P. Hu, Z. Q. Wang, X. X. Wu for useful discussions. This work was supported by the Photon and Quantum Basic Research Coordinated Development Program from MEXT, JSPS (KAKENHI Grants No. 25220707), the National Natural Science Foundation of China (No.11234014 and No.11504117), the Ministry of Science and Technology of China (2016YFA0401000, 2015CB921300), and the Chinese Academy of Sciences (XDB07000000). G. D. Gu is supported by the U.S. Department of Energy, Office of Basic Energy Sciences, Division of Materials Sciences and Engineering, under Contract No. DE-SC0012704.

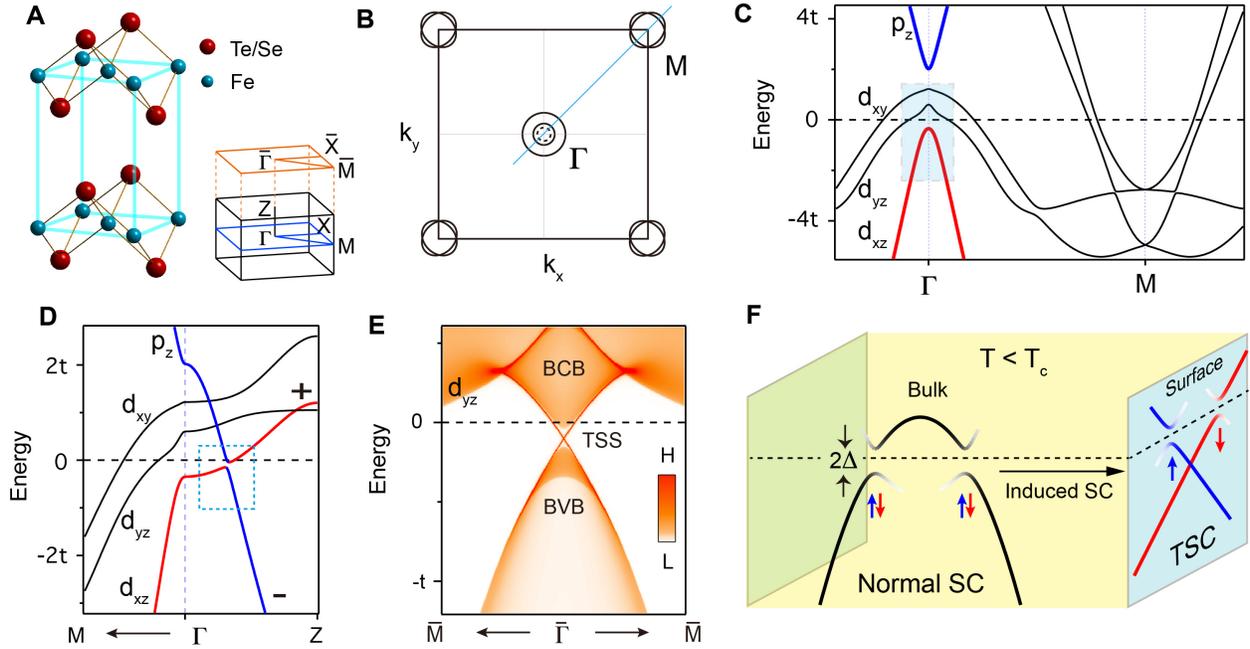

Fig.1 **Band structure and topological superconductivity of FeTe$_{0.5}$Se$_{0.5}$.** (**A**) Crystal structure of Fe(Te,Se), together with the 3D BZ and projected surface BZ. (**B**) In-plane BZ at $k_z = 0$. There are two hole-like FSs at Γ and two electron-like FSs at M. The dashed circle at Γ indicates the hole-like band just below $E_F$. (**C**) First-principles calculations on band structure along the ΓM direction, as indicated by the light blue line in (B). In the calculations, t = 100 meV, while t ~ 12 - 25 meV from experiments, largely depending on the bands (*27*). In this study, we focus on the small area around Γ with light blue background, where mainly d$_{xz}$ band is present. (**D**) First-principles calculations on band structure along ΓM and ΓZ. The dashed box shows the SOC gap of the inverted bands. (**E**) Band structure projected onto the (001) surface. The topological surface states (TSS) between the bulk valence band (BVB) and bulk conduction band (BCB) is clearly shown in this plot. (**F**) When the temperature decreases to below Tc, the bulk bands open s-wave SC gaps, which induce the surface states to be superconducting. The spin-helical surface states thus become topologically superconducting. (The side surface is shown for convenience.)

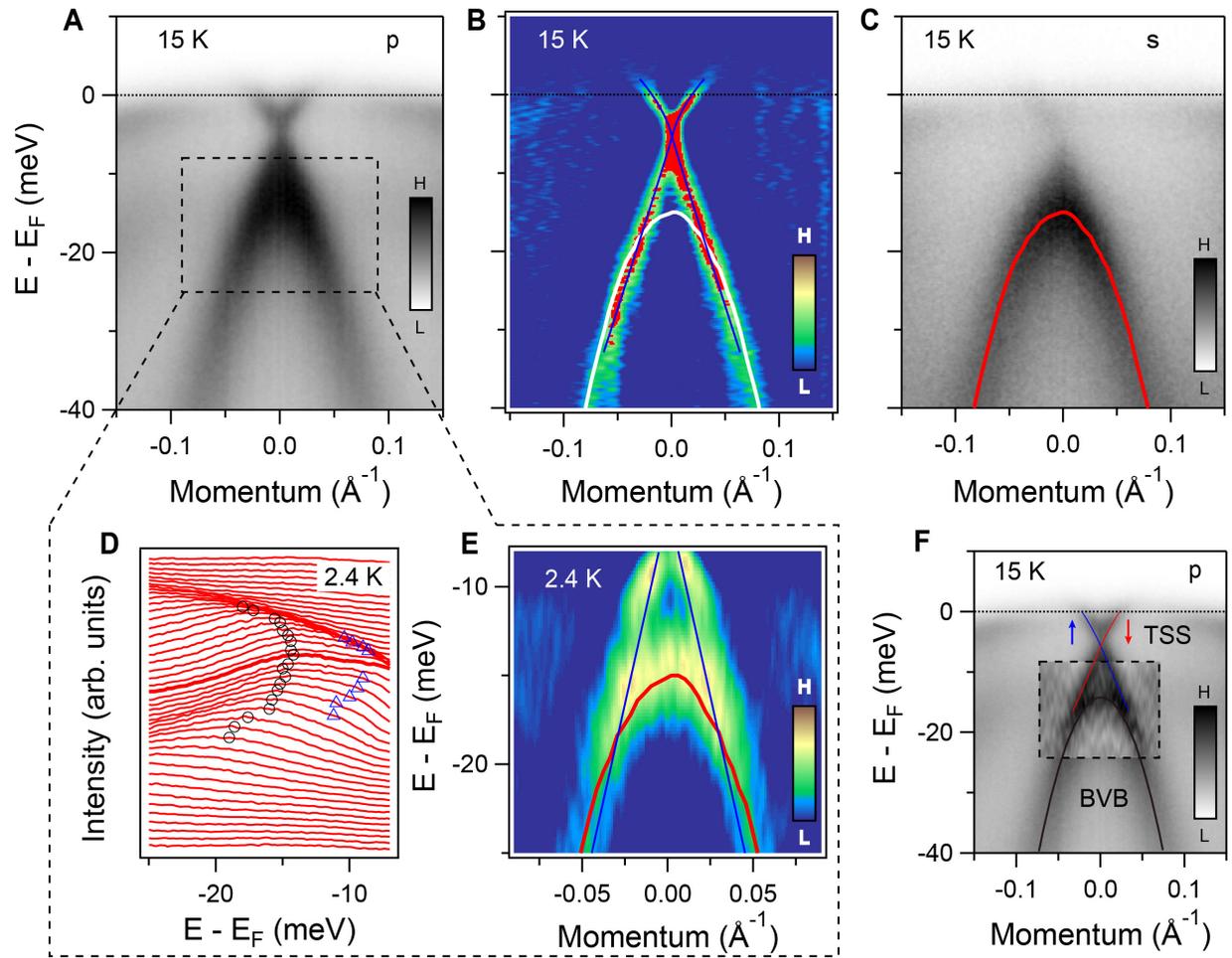

Fig.2 **Dirac-cone type surface band.** (**A**) Band dispersion along ΓM, recorded with a p-polarized 7-eV laser. (**B**) MDC curvature plot of (A), which enhances vertical bands (or vertical part of one band) but suppresses horizontal bands (or horizontal part of one band) (*25*). The red dots trace the points where the intensity of the MDC curvature exceeds a threshold, and the blue lines are guides to the band dispersion. The white line is the same as the red line in (C). (**C**) Same as (A), but recorded with s-polarized light. The red line comes from the Lorentzian fitting of the EDC peaks. (**D** and **E**) Zoom-in view of the dashed box area in (A). The data is recorded at 2.4 K to reduce the thermal broadening. (D) EDCs of the zoom-in area. The black and blue markers trace the EDC peaks. (E) EDC curvature plot of the zoom-in area. The blue lines are the same as the ones in (B), and the red line is the same as the one in (C). (**F**) Summary of the overall band structure. The background image is a mix of raw intensity and EDC curvature (the area in the dashed box). The bottom hole-like band is the bulk valence band, while the Dirac-cone type band is the surface band.

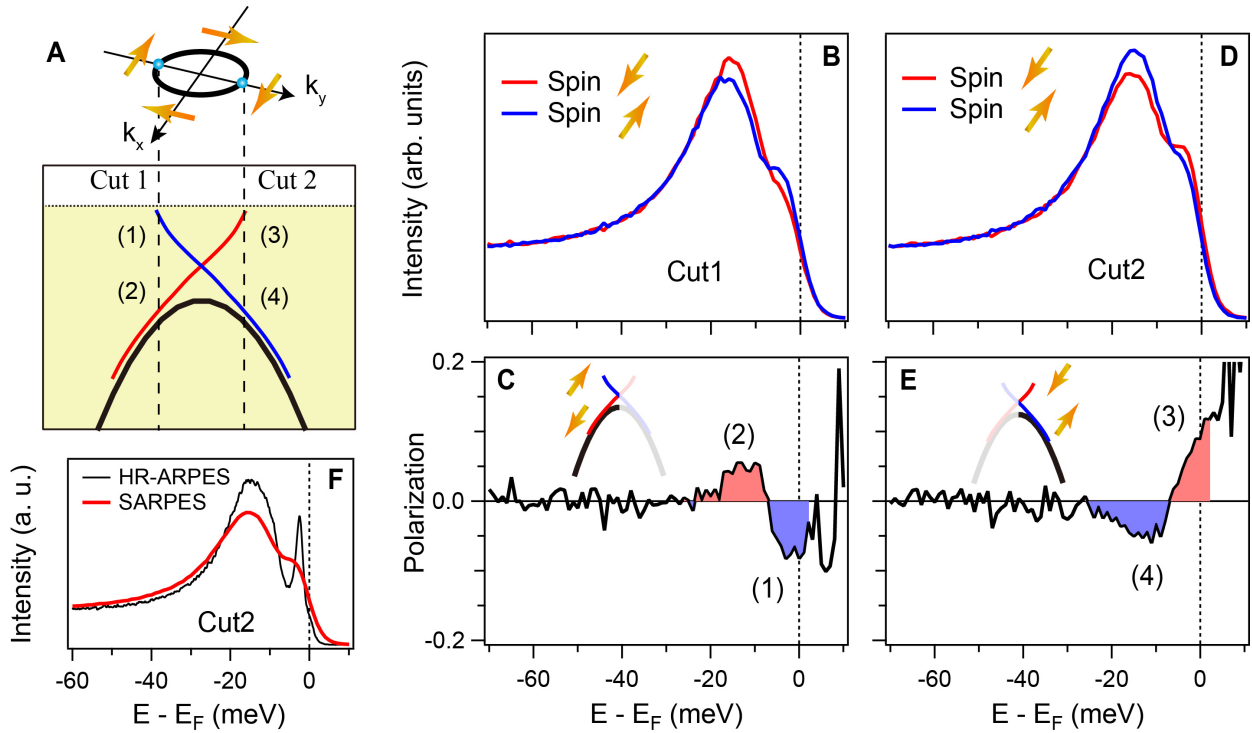

Fig.3 **Spin-helical texture of the surface band.** (**A**) Sketch of the spin-helical FS and the band structure along $k_y$, the sample ΓM direction. The EDCs at Cut1 and Cut2 were measured with spin-resolved ARPES. Note that the spin pattern shown in Reference (*12*) comes from the bottom surface. (**B**) Spin resolved EDCs at Cut1. (**C**) Spin polarization curve at Cut1. (**D** and **E**) Same as (B and C), but for EDCs at Cut2. The spin polarizations are consistent with the spin-helical texture illustrated in (A). (**F**) Comparison of the EDCs from SARPES measurement and HR-ARPES measurement. The large broadening in SARPES measurement could be partly responsible for the small spin polarization measured in (C) and (E).

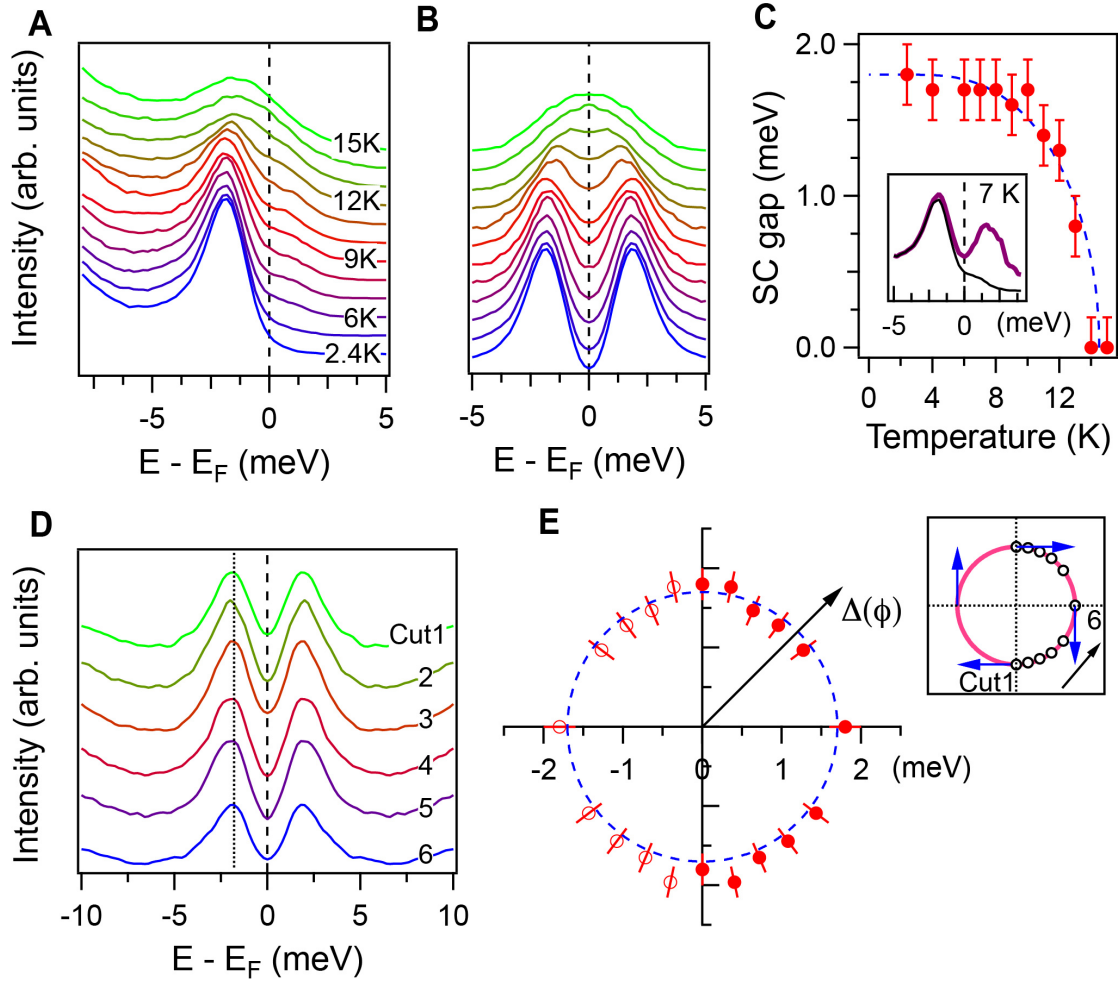

Fig.4 **s-wave SC gap of the surface band.** (**A**) Raw EDCs with different temperatures for a cut at the surface FS. The shoulders above $E_F$ are the sign of the SC Bogolubov quasiparticles. (**B**) Symmetrized EDCs of the curves shown in (A). (**C**) SC gap size as a function of temperature. The inset shows the raw EDC at 7K and the EDC after dividing the Fermi function, which clearly shows the Bogolubov quasiparticles above $E_F$. (**D**) Symmetrized EDCs at different Fermi wave vectors ($k_F$) recorded at 2.4 K. The $k_F$ positions of the cuts are indicated in (E). (**E**) The polar representation of the SC gap size. The hollow markers are a mirror of the solid markers. The panel on the right side shows the positions of different cuts on the surface FS.

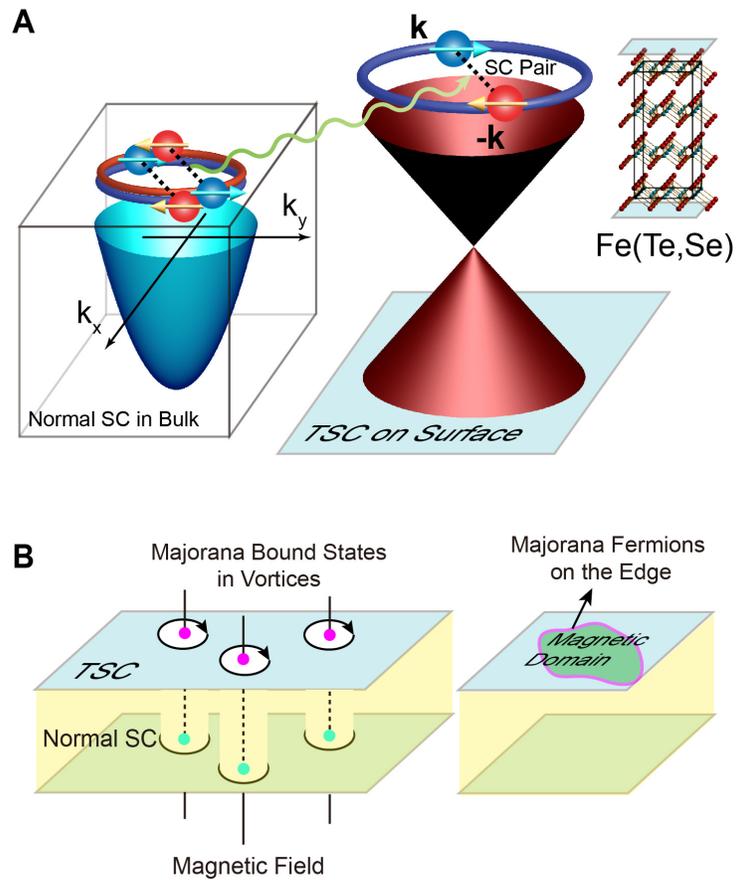

Fig.5 **Topological superconductivity and Majorana fermions on the surface.** (**A**) Topological superconductivity on the surface of FeTe$_{0.55}$Se$_{0.45}$. The electrons in the bulk are not spin-polarized, and the s-wave SC pairing is topologically trivial. The electrons on the surface are induced to form SC pairs by the bulk superconductivity. The superconductivity of the spin-helical surface states is topologically non-trivial. (**B**) Magnetic field creates vortices in FeTe$_{0.55}$Se$_{0.45}$, which behave as edges for the topological superconductivity on the surface. Thus, there will be Majorana bound states in the vortices. If there is a magnetic domain on the surface that kills superconductivity in the domain, there will be Majorana fermions along the boundary of the domain.